\documentclass{emulateapj}
\usepackage{apjfonts}
\usepackage{color}

\def\au{{\rm AU}}

\def\masyr{{\rm mas}\,{\rm yr}^{-1}}
\def\kpc{{\rm kpc}}
\def\mas{{\rm mas}}
\def\muas{\mu{\rm as}}

\def\max{{\rm max}}
\def\rel{{\rm rel}}

\def\e{{\rm E}}

\def\eqalign#1{\null\,\vcenter{\openup\jot
        \ialign{\strut\hfil$\displaystyle{##}$&$
        \displaystyle{{}##}$\hfil \crcr#1\crcr}}\,}

\lefthead{MR{\'O}Z ET AL.} 
\righthead{OGLE-2016-BLG-0596L\MakeLowercase{b}}

\begin{document}

\title{OGLE-2016-BLG-0596L\MakeLowercase{b}: High-Mass Planet From High-Magnification 
Pure-Survey Microlensing Event}

\author{
P. Mr{\'o}z$^{1}$,
C. Han$^{2,3}$,\\
and\\
A. Udalski$^{1}$,
R. Poleski$^{1,4}$,
J. Skowron$^{1}$,
M.\,K. Szyma{\'n}ski$^{1}$,
I. Soszy{\'n}ski$^{1}$,
P. Pietrukowicz$^{1}$,
S. Koz{\l}owski$^{1}$,
K. Ulaczyk$^{1,5}$,
{\L}. Wyrzykowski$^{1}$,
M. Pawlak$^{1}$\\
(OGLE group)\\
M. D. Albrow$^{6}$,
S.-M. Cha$^{7,8}$,
S.-J. Chung$^{7}$,
Y. K. Jung$^{9}$,
D.-J. Kim$^{7}$,
S.-L. Kim$^{7,10}$,
C.-U. Lee$^{7,10}$,
Y. Lee$^{7,8}$,
B.-G. Park$^{7,10}$,
R. W. Pogge$^{4}$,
Y.-H. Ryu$^{7}$,
I.-G. Shin$^{9}$,
J. C.\ Yee$^{9,11}$,
W. Zhu$^{4}$,
A. Gould$^{7,12,4}$\\
(KMTNet group)\\
}

\affil{$^{1}$  Warsaw University Observatory, Al.~Ujazdowskie~4, 00-478~Warszawa,Poland}
\affil{$^{2}$  Department of Physics, Chungbuk National University, Cheongju 361-763, Republic of Korea}
\affil{$^{3}$  Corresponding Author}
\affil{$^{5}$  Department of Physics, University of Warwick, Gibbet Hill Road, Coventry, CV4 7AL, UK}
\affil{$^{4}$  Department of Astronomy, Ohio State University, 140 W. 18th Ave., Columbus, OH  43210, USA}
\affil{$^{6}$  University of Canterbury, Department of Physics and Astronomy, Private Bag 4800, Christchurch 8020, New Zealand}
\affil{$^{7}$  Korea Astronomy and Space Science Institute, Daejon 305-348, Republic of Korea}
\affil{$^{8}$  School of Space Research, Kyung Hee University, Yongin 446-701, Republic of Korea}
\affil{$^{9}$  Harvard-Smithsonian Center for Astrophysics, 60 Garden St., Cambridge, MA 02138, USA}
\affil{$^{10}$ Korea University of Science and Technology, 217 Gajeong-ro, Yuseong-gu, Daejeon 34113, Korea}
\affil{$^{11}$ Sagan Fellow}
\affil{$^{12}$ Max-Planck-Institute for Astronomy, K\"onigstuhl 17, 69117 Heidelberg, Germany}

\begin{abstract}
We report the discovery of a high mass-ratio planet $q=0.012$, i.e., 13 times higher 
than the Jupiter/Sun ratio.  The host mass is not presently measured but can be determined 
or strongly constrained from adaptive optics imaging.  The planet was discovered in a small 
archival study of high-magnification events in pure-survey microlensing data, which was 
unbiased by the presence of anomalies.  The fact that it was previously unnoticed may 
indicate that more such planets lie in archival data and could be discovered by similar 
systematic study.  In order to understand the transition from predominantly survey+followup 
to predominately survey-only planet detections, we conduct the first analysis of these 
detections in the observational $(s,q)$ plane.  Here $s$ is projected separation in units 
of the Einstein radius.  We find some evidence that survey+followup is relatively more 
sensitive to planets near the Einstein ring, but that there is no statistical difference 
in sensitivity by mass ratio.
\end{abstract}


\section{Introduction}

For the first decade of microlens planet detections, beginning with 
OGLE-2003-BLG-235Lb \citep{ob03235}, the great majority of detections 
required a combination of survey and followup data.  This is a consequence 
of two effects. First, the survey coverage was generally too sparse to 
characterize the planetary anomalies in the detected events \citep{gouldloeb}.  
Second, thanks to aggressive alert capability, pioneered by the Optical 
Gravitational Lensing Experiment (OGLE) Early Warning System 
(EWS, \citealt{ews1,ews2}), it became possible to organize intensive 
followup of planet-sensitive events -- or even ongoing planetary anomalies 
-- and so obtain sufficient time resolution to detect and characterize planets.

However, as surveys have become more powerful over the past decade, they
have become increasingly capable of detecting planets without followup 
observations.  That is, making use of larger cameras, the surveys are able 
to monitor fairly wide areas at cadences of up to several times per hour.  
While still substantially lower than followup observations of the handful 
of events that were monitored by followup groups, this is still adequate 
to detect most planets (provided that the anomalies occur when the survey 
is observing).  Very simple reasoning given below, which is supported by 
detailed simulations \citep{Zhu:2014}, leads one to expect that the transition 
from survey+followup to survey-only mode implies a corresponding transition 
from planets detected primarily in high-magnification events via central 
and resonant caustics to planets primarily detected in lower magnification 
events via planetary caustics.

High-magnification events are intrinsically sensitive to planets because
they probe the so-called ``central caustic'' that lies close to (or overlays)
the position of the host \citep{griest98}.  Planets that are separated
from the hosts by substantially more (less) than the Einstein radius
generate one (two) other caustics that are typically much larger
than the central caustic and thus have a higher cross section for
anomalous deviations from a point-lens light curve due to a random
source trajectory.  However, for high-magnification events, the
source is by definition passing close to the host and hence close
to or over the central caustic.  For planet-host separations that
are comparable to the Einstein radius, the two sets of caustics merge
into a single (and larger) ``resonant caustic'', which is even more
likely to generate anomalous deviations of a high-magnification event.

For many years, the Microlensing Follow Up Network ($\mu$FUN) employed
a strategy based on this high planet sensitivity of high-magnification
events.  They made detailed analyses of alerts of ongoing events from 
the OGLE and the Microlensing Observations in Astrophysics (MOA) teams 
to predict high-magnification events and then mobilized followup observations 
over the predicted peak.  \citet{gould10} showed that $\mu$FUN was able to 
get substantial data over peak for about 50\% of all identified events with 
maximum magnification $A_\max>200$, but that its success rate dropped off 
dramatically at lower magnification, i.e., even for $100<A_\max<200$.  The 
reason for this drop off was fundamentally limited observing resources: 
there are twice as many events $A_\max>100$ compared to $A_\max>200$, and 
monitoring the full-width half-maximum requires twice as much observing time.  
Hence, observations grow quadratically with effective magnification cutoff.

By contrast, because planetary caustics are typically much larger than 
central caustics, most planets detected in survey-only mode are expected 
to be from anomalies generated by the former, which occur primarily in 
garden-variety (rather than high-mag) events \citep{Zhu:2014}.  For example, 
\citet{ob120406} detected a large planetary caustic in OGLE-2012-BLG-0406 
based purely upon OGLE data, while \citet{moabin1} detected one in MOA-bin-1 
based mostly on MOA data.  In the latter case it would have been completely 
impossible to discover the planet by survey+followup mode because the 
``primary event'' (due to the host) was so weak that it was never detected 
in the data.

Nevertheless, there has been a steady stream of survey-only detections
of planets in high-magnification events as well.  The first of these
was MOA-2007-BLG-192Lb, a magnification $A_\max> 200$ event, which required 
a combination of MOA and OGLE data \citep{mb07192}.  The first 
planet detected by combining three surveys (MOA, OGLE, Wise),
MOA-2011-BLG-322Lb,  was also via a central caustic, although in this case 
the caustic was very large so that the magnification did not have to be
extremely large $(A_\max\sim 20)$ \citep{mb11322}.  Similarly,
\citet{ob150954} detected a large central caustic due to the large planet
OGLE-2015-BLG-0954Lb despite modest peak magnification of the underlying
event $A_\max\sim 20$.  This case was notable because high-cadence
data from the Korea Microlensing Telescope Network (KMTNet) captured
the caustic entrance despite the extremely short source self-crossing
time, $t_*=16\,$min.
There also exist 2 planets MOA-2008-BLG-379Lb \citep{Suzuki2014} and
OGLE-2012-BLG-0724Lb \citep{Hirao2016} that were detected by the OGLE+MOA 
surveys through the high-magnification channel.

KMTNet is still in the process of testing its reduction pipeline.
Motivated by the above experience, the KMTNet team focused its
tests on high-magnification events identified as such on the OGLE
web page.  In addition to exposing the reduction algorithms to a wide
range of brightnesses, this testing has the added advantage that
there is a high probability to find planets.  Here we report
on the first planet found by these tests from among the first seven 
high-mag events that were examined: 
OGLE-2016-BLG-(0261,0353,0471,0528,0572,0596,0612).  These
events were chosen to have model point-lens magnifications $A>20$
and modeled peak times $2457439<t_0<2457492$.  The lower limit
was set by the beginning of the KMTNet observing season and the upper
limit was the time of the last OGLE update when the seven events were
selected.

\section{Observations}

On 2016 April 8 UT 12:15 (HJD$^\prime=$ HJD$-2450000=7487.0$), OGLE
alerted the community to a new microlensing event OGLE-2016-BLG-0596
based on observations with the 1.4 deg$^2$ camera on its 1.3m Warsaw 
Telescope at the Las Campanas Observatory in Chile \citep{ogleiv}
using its EWS real-time event detection software \citep{ews1,ews2}.  
Most observations were in $I$ band, but with some $V$ band observations 
that are, in general, taken for source characterization.  These $V$-band 
data are not used in the modeling.
At equatorial coordinates 
$(17^{\rm h} 51^{\rm m} 12^{\rm s}\hskip-2pt .81, -30^\circ 50' 59''\hskip-2pt.4)$ 
and Galactic coordinates $(-1.01^\circ,-2.03^\circ)$, 
this event lies in OGLE field BLG534, with an observing cadence during
the period of the anomaly of
roughly 0.4 per hour\footnote{OGLE cadences were significantly
adjusted at the time of the peak and planetary anomaly of this event,
due to the {\it Kepler} K2 Campaign 9 microlensing campaign.  The five
fields covering the K2 field were observed 3 times per hour, while other
fields (including BLG534) were observed somewhat less frequently (very 
roughly 2/3) compared to their usual rates.}.

KMTNet employs three $4.0\,\rm deg^2$ cameras mounted on 1.6m telescopes 
at CTIO/Chile, SAAO/South Africa, and SSO/Australia \citep{kmtnet}.
In 2015 KMTNet had concentrated observations on 4 fields. However,
in 2016, this strategy was radically modified to cover (12, 40, 80)
$\rm deg^2$ at cadences of $(4,\ \geq 1,\ \geq 0.4)\,\rm hr^{-1}$.
For the three highest-cadence fields, KMTNet observations are alternately
offset by about $6^\prime$ in order to ensure coverage of events
in gaps between chips.  As a result, OGLE-2016-BLG-0596 lies in
two slightly offset fields BLG01 and BLG41, which are each observed
at a cadence 2 per hour\footnote{Like OGLE, KMTNet also adjusted its
schedule for the K2 campaign, but in a different way.  First, CTIO
observations were not adjusted.  Second, KMTNet only began ``K2 mode''
on 2016 April 23.  This was after the event peak and caustic
entrance but before the exit.  Therefore, in particular, the caustic
exit observations from SAAO were at the lower cadence (reduced by a factor
0.75)}. KMTNet observes primarily in $I$ band, but 1/11 observations from 
CTIO and 1/21 observations from SAAO are in
$V$-band.

Reductions of the primary data were made using difference image analysis 
(DIA) \citep{alard98}.  However, due to issues discussed in Section 3,
special variants of DIA were developed specifically for this event.  
See below.  KMT CTIO $V$ and $I$ images were, in addition, reduced using 
DoPHOT \citep{dophot}, solely for the purpose of determining the source color.

\begin{figure}
\includegraphics[width=\columnwidth]{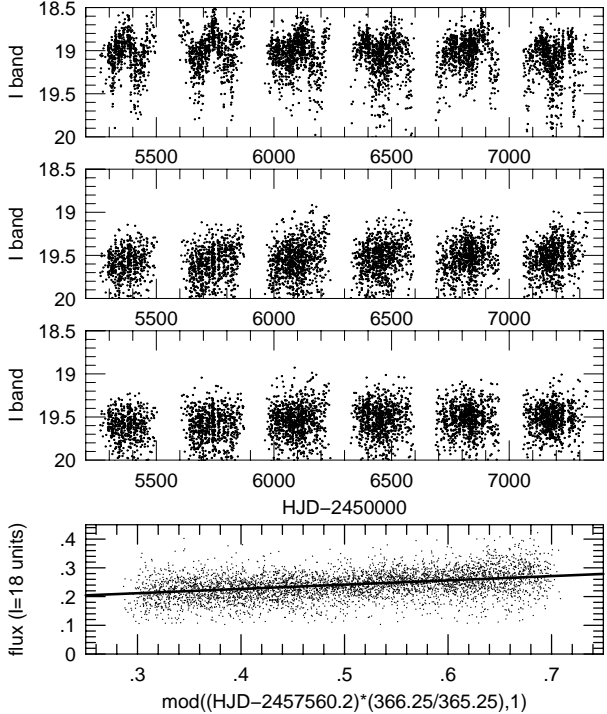}
\caption{ Correction of light curve variability.  Top panel shows
OGLE online reductions.  Variability is roughly periodic, $P\simeq 126.5$ 
days, and semi-amplitude of $\sim 8.7\%$ of baseline flux, but 57\% of
source flux.  Second panel shows result from simultaneous fit to
(1) microlensed source, (2) bright variable at $1.5^{\prime\prime}$,
(3) nearby blended star.  Periodic variability is removed but annual trend
remains.  Mean flux drops by 0.5 mag due to fitting out third star.
Bottom panel shows measured flux (from second panel) but as a function
of sidereal time.  This is well fit by a straight line. Note that
the full range of this fit to the variation is a factor $\sim 1.5$ 
larger than the source flux.  Final photometry (third panel) is obtained 
by subtracting this straight-line fit from all flux measurements (not 
just at baseline).  }
\label{fig:one}
\end{figure}

\section{Light Curve Variability}

\subsection{Evidence of Variability}

The OGLE-2016-BLG-0596 light curve shows clear variability over the
course of 6 seasons of OGLE-IV data prior to 2016.  This variability
is roughly consistent with being sinusoidal at period $P=126.5\,$ days. 
See Figure~\ref{fig:one}.  While the semi-amplitude of the variability is 
only $8.7\%$ of the baseline flux $(I_{\rm base}\sim 19)$, this semi-amplitude 
turns out to be roughly equal to the source flux derived from the model.

\subsection{Importance of Variability}

Assuming (as proves to be the case) that it is not the source itself 
that is variable, such low level variability cannot significantly impact 
characterization of the anomalous features of the lightcurve because they 
are at relatively high magnification and take place on much shorter timescales.  
However, if not properly accounted for, such variability can seriously impact 
the estimate of the source flux and, as a direct consequence of this, the 
Einstein timescale
\begin{equation}
t_\e = {\theta_\e\over\mu};
\qquad
\theta_\e^2\equiv \kappa M \pi_\rel;
\qquad
\kappa\equiv {4 G\over c^2\au}\simeq 8.14{{\rm mas}\over M_\odot}.
\label{eqn:tedef}
\end{equation}
Here $\theta_\e$ is the angular Einstein radius,
$\pi_\rel = \au(D_L^{-1}-D_S^{-1})$ is the lens-source relative parallax, 
and $\mu$ is the lens-source relative proper motion in the Earth frame.
Errors in these quantities would propagate into the estimates
of the planet-star mass ratio $q$, the Einstein radius $\theta_\e$,
and the proper motion $\mu$, all of which are important for assessing
the physical implications of the detection.

The reason that $t_\e$ is potentially impacted by unmodeled variability
is that it is determined primarily from the wings of the light curve
where the amplitude of the amplification of flux is comparable to that of
the variability.  Hence, it is important to track down the source
of this variability and correct for it to the extent possible.  See, e.g.,
\citet{mb11293}.

\subsection{Removal of Variability I: Variations of Neighbor}

In principle, tracking down such a low level of variability in such
a crowded field could have been very difficult.  However, in the
present case, it turns out to be due to a star $1.5^{\prime\prime}$
to the southeast, which is quite bright $I\sim 14.5$ and shows variability
with the same period and phase.  Within the framework of standard DIA,
it is natural that this variable should impact the microlensing light curve
because the difference image contains residuals from the variable that
overlap the point spread function (PSF) of the microlensed star.  Hence,
when the difference image is dot-multiplied by the PSF to estimate the 
flux, it includes a contribution from the residual flux of the variable.

It is straightforward to simultaneously fit for two (or $n$)
variables with possibly overlapping PSFs.  After constructing difference
images in the standard way, one simply generalizes the normal procedure by
calculating the $n$ flux-difference values
\begin{equation}
F_i = \sum_{j=1}^n c_{ij} d_j;
\qquad c\equiv b^{-1},
\label{eqn:multifit1}
\end{equation}
where
\begin{equation}
b_{ij} = \sum_k {P_{i,k}P_{j,k}\over \sigma_k^2};
\qquad
d_i = \sum_k {P_{i,k}f_k\over \sigma_k^2},
\label{eqn:multifit2}
\end{equation}
$P_{i,k}$ is the (unit normalized: $\sum_k P_{i,k}=1$) amplitude of the 
$i$th PSF in the $k$th pixel and  $(f_k,\sigma_k)$ are the value and error 
of the difference flux in the $k$th pixel.

We use a variant of this formalism to reduce the OGLE data with
$n=3$ stars, including the microlensed source, the bright variable,
and one other very nearby (but non-variable) blended star.  The result 
is shown in the second panel of Figure~\ref{fig:one}.  First note that 
the fluxes have decreased by about 0.5 mag because the non-variable 
neighboring blend (third star in the fit) has been removed from the 
baseline flux.  The semi-periodic variations are removed.  However 
there remains an annual trend.

\subsection{Removal of Variability II: Annual Variations}

The bottom panel of Figure~\ref{fig:one} shows that this annual
trend is due to variations with sidereal time, almost certainly due
to the impact of the bright red neighbor (even if it were constant)
via differential refraction.  We fit this variation to polynomials
of order $n$, but find that there is no significant improvement beyond
$n=1$, for which $f(t) = 0.2415 + 0.1490(t-0.5)$, where $f$ is flux
in units of $I=18$.  Note that the variation from 0.3 to 0.7 (on the figure)
is $0.4\times 0.149\sim 0.06$ which is 1.5 times larger than the source
flux derived below.

The third panel shows the results of applying this sidereal-time correction.
As expected the annual trend is gone.  We apply this flux correction
to all data, not just the baseline data shown in this figure.  We find
(as expected) that this corrects the slope of the rising part of the
light curve, which indeed impacts the estimate of $t_\e$, though
by less than 10\%.  Note that while, in most cases, it is possible
to derive an accurate estimate of the photometric error bars
from those reported by the photometric pipeline \citep{skowron16},
in this case we did not apply this simple prescription because
the data were reduced using a special customized pipeline.

\subsection{Correction of KMTNet Data}

We apply the same formalism given by Equations~(\ref{eqn:multifit1}) and
(\ref{eqn:multifit2}) to the KMTNet data, but with only two stars, i.e.,
the microlensed source and the neighboring variable.  Note that this 
difference between OGLE and KMTNet reductions plays no role in the final 
result because the third star incorporated into the OGLE fit is not variable, 
and the KMTNet flux scale is ultimately aligned to OGLE through the microlens 
fit.  Thus, in particular, we retain the advantage of resolving out this blend, 
thus placing better limits on flux from the lens.

We note that it is difficult to correct for the annual variation in the KMTNet data. 
Reliable measurement of the annual variation would require baseline data, 
which do not exist because the photometry system was not the same for the 
whole of 2015 compared to 2016.  In principle, we could have applied the 
OGLE-based correction to KMTNet data, but this type of correction is 
observatory-specific and this would not have been a reliable approach and 
could easily cause more problem than what was being corrected for. 
From checking the impact of only correcting the OGLE data for 
variable-contamination (but not annual variation), however, we found essentially no 
change. The scatter (hence renormalized error bars) are very slightly smaller, 
but no change in parameters. The same would be case for KMTNet data.

\begin{figure}
\includegraphics[width=\columnwidth]{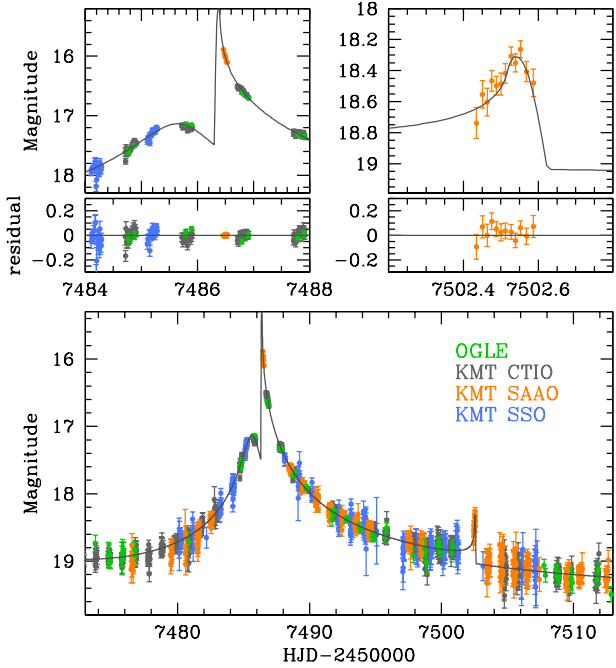}
\caption{ Light curve and geometry of OGLE-2016-BLG-0596.
The event is primarily characterized by a strong caustic entrance
at HJD$^\prime\sim 7486$ superposed on an otherwise slightly
asymmetric point-lens-like light curve.  There is a weak caustic
exit at HJD$^\prime\sim 7502.5$ which is well covered by KMT SAAO data.
This morphology, together with the $\sim 16$ day interval from caustic
entrance to exit, is indicative of a resonant caustic (top panel) due
to a high-mass planet or low-mass brown dwarf.  }
\label{fig:two}
\end{figure}

\subsection{Guideline for Assessing the Need of Multi-star Fitting}

The formalism introduced in Section 3.3
can also be used to gain intuition about the impact
of uncorrected variability, which can then be used to assess whether
such corrections are necessary in specific cases.

First note that for $n=1$, Equations~(\ref{eqn:multifit1}) and 
(\ref{eqn:multifit2}) reduce to the standard formula:
\begin{equation}
F_1 = {d_1\over b_{11}};
\qquad
b_{11} = \sum_k {P_{1,k}^2\over \sigma_k^2};
\qquad
d_1 = \sum_k {P_{1,k}f_k\over \sigma_k^2}
\label{eqn:singlefit}
\end{equation}

Equation~(\ref{eqn:singlefit}) then allows us to express the properly 
corrected ``true'' photometry in terms of the ``naive'' single-source
photometry that ignores neighbors.  We first ``infer'' the value,
$d_{i,\rm inferred} = b_{ii}*F_{i,\rm naive}$, which then yields,
\begin{equation}
F_{i,\rm true} = \sum_{j=1}^2 c_{ij}d_{j,\rm inferred}=
{F_{i,\rm naive} - (b_{12}/b_{ii})F_{(3-i),\rm naive}\over
1 - b_{12}^2/(b_{11}b_{22})}.
\label{eqn:backout}
\end{equation}

Hence, the correction is governed by the ratio of the PSF overlap integral
$b_{12}$ to the integral of the PSF squared, $b_{11}$.  We can evaluate 
this explicitly for the special case of a Gaussian PSF and below-sky sources
($\sigma_k =$const)
\begin{equation}
{b_{12}\over b_{11}} = 4^{-(\Delta \theta/{\rm FWHM})^2}
\qquad
(\rm below-sky\ Gaussian),
\label{eqn:gaussbelow}
\end{equation}
where $\Delta\theta$ is the separation between the two sources.
If the two sources are reasonably well separated, $\Delta\theta\ga$FWHM,
and (as in the present case) the target (1) is below sky while the
contaminating variable (2) is well above sky, then the effect is
roughly half of that given by Equation~(\ref{eqn:gaussbelow}).  This
is because the squared PSF integral is basically unaffected while
the half of the contribution to the overlap integral that is closer
to the contaminant is heavily suppressed by the higher flux errors
per pixel.  We close by re-emphasizing that these order-or-magnitude
estimates are not used in the present analysis but are intended
as guidance for future cases.

\begin{figure}
\includegraphics[width=\columnwidth]{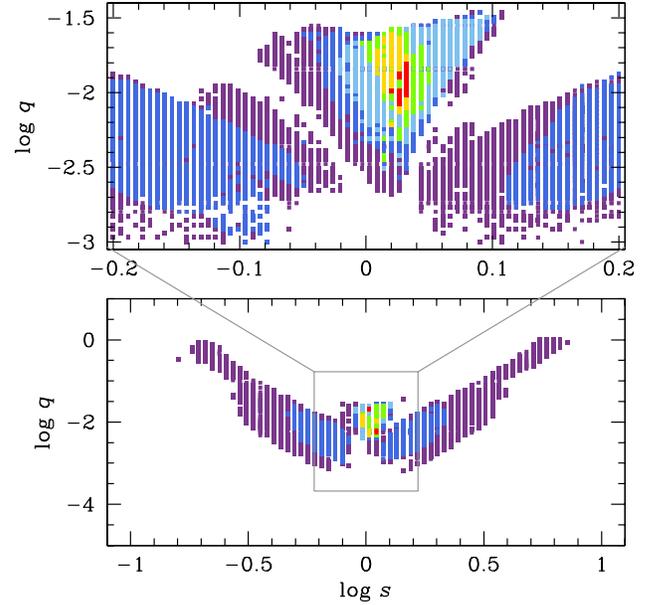}
\caption{ 
$\Delta\chi^2$ map of the MCMC chain in the 
$\log s$--$\log q$ parameter space obtained from the preliminary grid search.
The lower panel shows the entire range where the grid search is conducted.
The upper panel show the enlarged view around the best-fit solution.
Color coding represents MCMC points with $\leq 1n\sigma$ (red), $2n\sigma$ (yellow), 
$3n\sigma$ (green), $4n\sigma$ (cyan), $5n\sigma$ (blue), and $6n\sigma$ (magenta) 
of the best-fit and $n=10$.
}
\label{fig:three}
\end{figure}

\section{Light Curve Analysis}

The lightcurve, presented in Figure~\ref{fig:two},
has two principal features: a strong caustic entrance
near peak at HJD$^\prime = 7486.4$ and a weak caustic exit 
at HJD$^\prime = 7502.6$.    Apart from these caustic crossings, the morphology
is that of a slightly distorted point-lens event.  This morphology 
points to a binary lens with very unequal mass ratio $q\ll 1$, i.e., 
in the brown-dwarf or planetary regime.  The long duration of the caustic
(16 days) then points to a resonant caustic, and so projected separation
(in units of $\theta_\e$) of $s\sim 1$. 
A thorough search of the parameter space 
spanning $-1.0\leq \log s \leq 1.0$ and $-5.0\leq \log q \leq 1.0$ 
leads to only one viable solution, 
which confirms the above naive reasoning.
The uniqueness of the solution is shown in Figure~\ref{fig:three}, where 
we present the $\Delta\chi^2$ map of the MCMC chain in the $\log s$--$\log q$ 
parameter space obtained from the preliminary grid search.  In fact, initial 
modeling based on data taken up through  HJD$^\prime = 7500.8$ (so, before 
the caustic exit) already led to essentially this same solution (although 
the predicted caustic exit was 2.6 days later than the one subsequently observed).

The model is described by seven parameters. These include three that 
are analogous to a point-lens event $(t_0,u_0,t_\e)$, i.e., the
time of closest approach to the center of magnification, the impact
parameter normalized to $\theta_\e$ and the Einstein crossing time;
three to describe the binary companion $(s,q,\alpha)$ where $\alpha$
is the angle of binary axis relative to the source trajectory; and
$\rho\equiv \theta_*/\theta_\e$, where $\theta_*$ is the angular radius
of the source.

\begin{figure}
\includegraphics[width=\columnwidth]{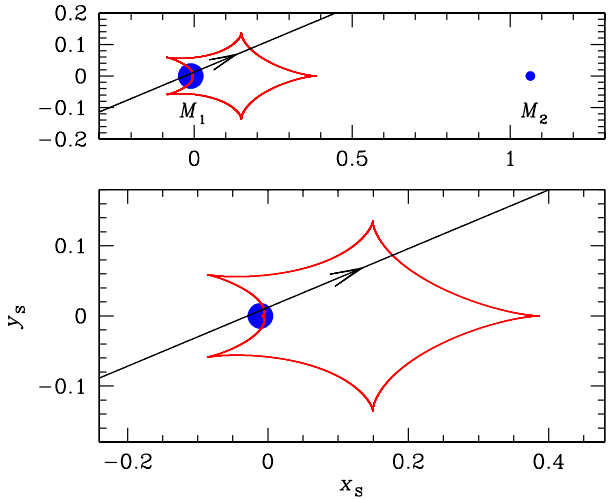}
\caption{ Caustic geometry of OGLE-2016-BLG-0596.  Top panel shows
the causic (red) with the positions of the host (left) and planet
(right) represented as blue circles.  Zoom in lower panel shows
that the source passed close to, but did not cross (because
$u_0>\rho$), the small central cusp.  In this region, the caustic
is very strong, accounting for the sharp jump at HJD$^\prime\sim 7486.4$.
On the other hand the caustic exit to the right is very weak, which
accounts for the smallness of the corresponding bump in the light curve
at HJD$^\prime\sim 7502.6$. }
\label{fig:four}
\end{figure}

\begin{deluxetable}{ccrr}
\tablecaption{Microlensing parameters for OGLE-2016-BLG-0596\label{table:one}}
\tablewidth{0pt}
\tablehead{
\multicolumn{1}{c}{Parameter}     &
\multicolumn{1}{c}{Unit}         &
\multicolumn{1}{c}{best}         & 
\multicolumn{1}{c}{error} 
}
\startdata
$t_0$               & day         & 7486.464 & 0.010  \\
$u_0$               & $10^{-2}$   &    1.112 & 0.031  \\
$t_{\rm E}$         & day         &    81.694 & 2.195 \\
$s$                 &             &     1.075 & 0.003 \\
$q$                 & $10^{-2}$   &     1.168 & 0.040 \\
$\alpha$            & radian      &     5.886 & 0.009 \\
$\rho$              & $10^{-2}$   &     0.060 & 0.008 \\
$I_s$               &             &    21.510 & 0.028 \\
$I_b$               &             &    19.739 & 0.028 
\enddata
\end{deluxetable}

The best fit parameters and errors (determined from a Markov Chain) are 
given in Table~\ref{table:one}.  
We present the model light curve superposed on the data points in 
Figure~\ref{fig:two} and the lens geometry is shown in Figure~\ref{fig:four}.
We also fit the lightcurve for the microlens parallax effect, but found no 
improvement.

We note that compared to other planetary and binary events with well-covered
caustic crossings, the parameter $\rho=(6.0\pm 0.8)\times 10^{-4}$ 
(and the parameter combination $t_* = \rho t_\e = 0.049 \pm 0.007$) have
relatively large errors.  These parameters are usually better measured 
because caustic crossings tend to be bright (since the caustic itself 
is a contour of formally infinite magnification), which means that the 
photometry over the caustic crossing is relatively precise.  Since $t_*$ 
depends almost entirely on the duration of this crossing, with only weak 
dependence on other model parameters, it can then be determined quite precisely.

In the present case, however, the first crossing was entirely missed
simply because it was not visible from any of the five survey telescopes
currently in operation (OGLE, MOA, and three from KMTNet: CTIO, SAAO, SSO).
The caustic exit was captured by KMTNet SAAO, with 12 points taken over 
3.63 hours (i.e., 20 minute cadence).  However, since this caustic was 
quite weak, peaking at $I\sim 18.1$, the photometry has much larger errors 
than the SAAO photometry near peak.  See upper two panels of Figure~\ref{fig:two}.

Adopting a more glass-half-full orientation, we should assess the
prior probability that either of the two caustic crossings would have
been adequately observed to measure $\rho$.  Considering the 20 days
between HJD$^\prime$ 7485 and 7505, the three KMTNet observatories
each took at least two points on 13 nights, with total durations,
(2.11,2.76,2.52) days for SSO, SAAO, and CTIO, i.e., a total 7.39 days.
Essentially all of these 39 intervals had approximately continuous coverage.
We estimate that the probability that $\rho$ can be measured is the same
as the probability that the caustic peak is covered, which may be
slightly too conservative.  Under this assumption, the probability
that the caustics would be observed are each 37\%, so that the probability
that at least one would be observed is $1-(1-0.37)^2=46\%$.
Of course, since the midpoint of the two caustic crossings was 16 April,
this probability is adversely affected by the shortness of the bulge
observing window relative to microlensing ``high season'' (21 May -- 21 July).
At that time the observing window is roughly 2.5 hours longer, and so
(assuming comparable weather conditions), the probability for each
crossing would be 52\% and the probability for at least one would be 77\%.
Nevertheless, the mid-April values may be considered as a proxy for
the microlensing season as a whole.


\begin{figure}
\includegraphics[width=\columnwidth]{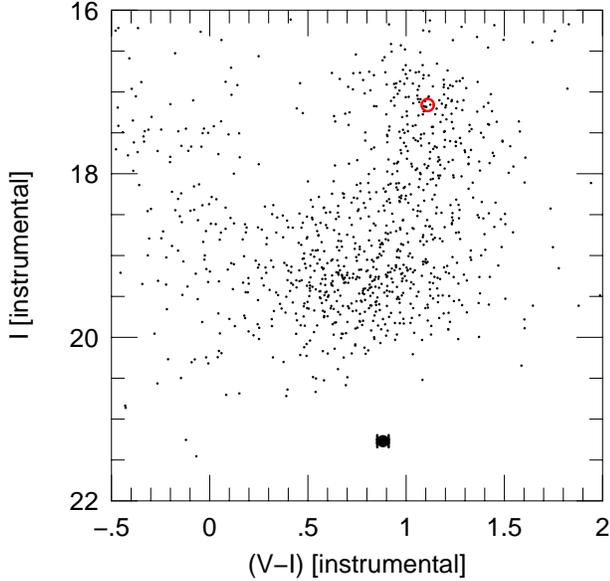}
\caption{ Instrumental CMD for $100^{\prime\prime}$ square around
OGLE-2016-BLG-0596 using KMTNet CTIO data.  
The instrumental source color is measured from model-independent regression 
and the instrumental magnitude is measured from the fit of the $I$-band 
data to the model light curve.
By measuring the offset of this source position from that of the red clump 
(red), one can determine the angular source radius $\theta_*$, using standard 
techniques \citep{ob03262}, as described in the text.  }
\label{fig:five}
\end{figure}

\section{Physical Parameters}

We use KMTNet CTIO DoPHOT reductions to construct an instrumental color magnitude 
diagram (CMD) that is presented in Figure~\ref{fig:five}.\footnote{
Correction of the DoPHOT data for variation is not done, but this would
have little effect on the result. The variable is extraordinarily red, 
$\sim 1.2$ magnitude redder than the clump, whereas the source is $\sim 0.2$ 
magnitude bluer than the clump. Hence, by a naive estimate, the variations 
would be fractionally smaller by a factor 4. The full amplitude of these 
variations in $I$ band is of order the source flux, whereas the color measurement 
is made when the source is magnified 60 to 100 times.  The color measurement is 
differential over short timescales of a few days, whereas the period is a large 
fraction of a year. Combining these very small factors, we expect the color 
measurement to be impacted at the level 
$(1/4)\times (1/80)\times (3/(126/\pi)) \sim 2\times 10^{-4}$. It is general 
practice to ignore such small errors, which in this case are more than hundred 
times smaller than the measurement error.
We also note that the dependence of the color measurement on the choice of the $V$-band 
data set (OGLE or KMTNet) is small considering that the offset from the clump has an 
accuracy of 0.03 magnitude whereas the precision of the color measurement is 0.05 magnitude. 
Furthermore, the SAAO $V$-band data are taken for redundancy, primarily in a case 
there is no CTIO data due to bad weather when the event is well magnified or 
for very short, highly magnified events that peak of South Africa. 
}  
We find the instrumental source color from model-independent regression and 
the instrumental source magnitude by fitting the $I$ band light curve to the 
model.  We then find the offset
from the clump $\Delta[(V-I),I]=(-0.23,4.06)\pm (0.03,0.10)$, where the
error in the color offset is dominated by the regression measurement while
the error in the magnitude offset is dominated by fitting for the clump
centroid.  We then adopt $[(V-I),I]_{0,\rm clump}=(1.06,14.49)$ 
\citep{bensby13,nataf13} to obtain $[(V-I),I]_{0,s}=(0.83,18.55)$.
Then using standard techniques \citep{ob03262}, we convert from $V/I$
to $V/K$ using the \citet{bb88} $VIK$ color-color relations and then use
the \citet{kervella04} color/surface-brightness relations to derive
\begin{equation}
\eqalign{
\theta_* = 0.690\pm 0.065\,\ \muas; \cr 
\theta_\e = {\theta_*\over\rho} = 1.15\pm 0.18\ \mas; \cr 
\mu = {\theta_\e\over t_\e} = 5.1\pm 0.8\,\ \masyr .\cr
}
\label{eqn:thetastar}
\end{equation}
The error in $\theta_*$ is dominated by the uncertainties
in transforming from color to surface brightness (8\%), with a significant
contribution from the error in $I_s$ (5\%).  The fractional errors in
$\theta_\e$ and $\mu$ are substantially larger than in $\theta_*$ due to
the relatively large error in $\rho$.  See Section 4.

The relatively large value of $\theta_\e$ almost certainly implies that
the lens lies in the Galactic disk since the lens-source relative parallax is
\begin{equation}
\pi_\rel = {\theta_\e^2\over\kappa M} = (0.16\pm 0.05 \mas)
\biggl({M\over M_\odot}\biggr)^{-1}.
\label{eqn:mpirel}
\end{equation}
That is, only if the lens were substantially heavier than $1\,M_\odot$
could it be in the bulge ($\pi_\rel\la 0.03$).  However, first, there
are almost no such massive stars in the bulge and second, its light would
then exceed the blended light $(I_b\sim 19.7)$, even allowing for the
$A_I=2.96$ extinction toward this line of sight \citep{nataf13}.  The
only exception to this line of reasoning would be if the lens were a black
hole.

\begin{figure}
\includegraphics[width=\columnwidth]{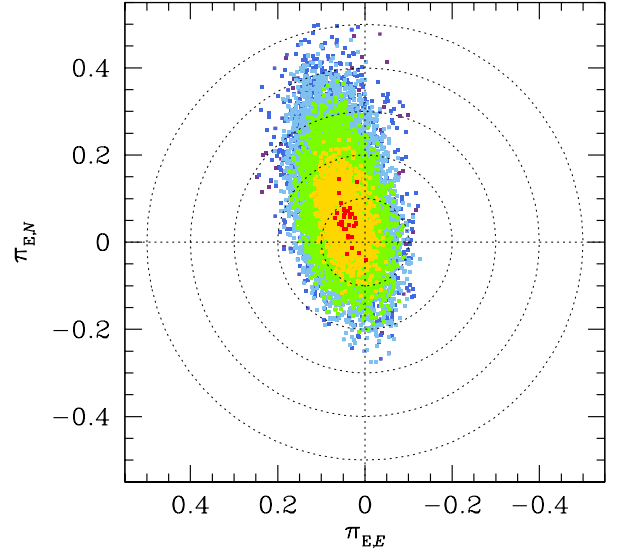}
\caption{
$\Delta\chi^2$ map of the MCMC chain in the $\pi_{{\rm E},E}$--$\pi_{{\rm E},N}$ 
parameter space. Color coding represents points in the MCMC chain with 
$\leq 1\sigma$ (red), $2\sigma$ (yellow), $3\sigma$ (green), $4\sigma$ (cyan), 
and $5\sigma$ (blue) of the best fit. The dotted circles represent 
the boundaries of $\pi_{\rm E}=0.1$, 0.2, 0.3, 0.4, and 0.5.
}
\label{fig:six}
\end{figure}

Although the model considering parallax effects does not have improvement 
compared to the non-parallax model, non-detection of $\pi_{\rm E}$ can 
give constraints on the mass and distance. In Figure~\ref{fig:six}, we present 
the $\Delta\chi^2$ map of the MCMC chain in the $\pi_{{\rm E},E}$--$\pi_{E,{\rm N}}$ 
parameter space obtained from the modeling considering both the lens orbital motion 
and the microlens parallax effect. The upper limit of the microlens parallax 
as measured 3$\sigma$ level is $\pi_{\rm E} \lesssim 0.4$. This gives the lower 
limits of the mass and distance of $M\gtrsim 0.35\ M_\odot$ and $D_{\rm L}\gtrsim 1.7$ kpc.

\section{Resolving the Nature of the Planet}

The most notable characteristic of OGLE-2016-BLG-0596 is its high mass ratio
$q=0.0117\pm 0.0004$, implying that the mass is 
$m_p = 12.2\,M_{\rm jup} (M/M_\odot)$.  Hence, if the host is one solar mass,
this planet would be just below the the deuterium-burning limit (usually
regarded as the planet/brown-dwarf boundary).  While the host could
in principle have arbitrarily low mass (and so, by Equation~(\ref{eqn:mpirel}),
be arbitrarily close), distances closer than $D_L\la 1\,\kpc$ are strongly
disfavored by the relatively low proper motion,
the parallax constraint, and 
the paucity of nearby lenses.  At this limiting distance, and so 
$M =\theta_\e^2/\kappa\pi_\rel \sim 0.18\,M_\odot$, the planet would still be
$m_p\sim 2\,M_{\rm jup}$, i.e., quite massive for such a low-mass host.
Hence, regardless of the host mass, this is a fairly extreme
system.

To distinguish among these interesting possibilities will require
measuring (or strongly constraining) the host mass.  This can be
accomplished with high resolution imaging, either using the {\it Hubble
Space Telescope (HST)} or ground-based adaptive optics (AO) imaging
on an 8m class telescope.  An advantage of {\it HST} is that it can
observe in the $I$ band for which the source flux is directly measured
from the event.  Hence, the source light can be most reliably separated
from the blended light in $I$.  In contrast to many previous cases, there
are no $H$ band observations during the event, so ground-based AO observations
(which must be in the infrared) cannot be directly compared to an
event-derived source flux.  Nevertheless, it is probably possible to
transform from $V/I$ light-curve measurement to $H_s$ source flux
with a precision of 0.2 mag, using a $VIH$ color-color diagram.

For definiteness, we will assume that the lens can be reliably detected
from {\it HST} or AO observations provided that the flux is at least half
that of the source, i.e., $I_L<22.3$.  For example, if the lens were
an 
 $M=0.5\,M_\odot$ 
early 
M dwarf 
(so $D_L\sim 2.2\,\kpc$), then it would have 
 $I_L \sim 18.8 + A_I\leq 21.8$ 
(and brighter if, as is almost certainly
the case, a substantial fraction of the dust is beyond 
2.2 kpc).
Thus, there is a good chance that AO or {\it HST} observations could
detect the lens, and even if this failed, the observations would strongly
constrain the host to be of very low mass.

\begin{figure}
\includegraphics[width=\columnwidth]{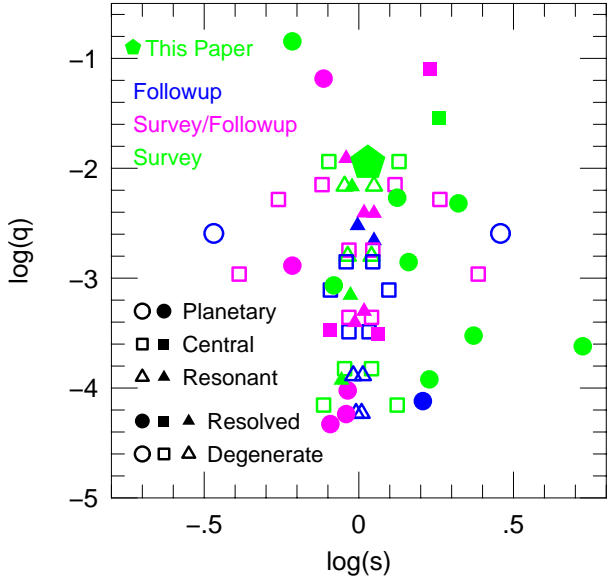}
\caption{
Log-log plot of planet-star mass ratio $q$ versus separation
(normalized to $\theta_\e$) $s$ for 44 previously published or submitted
planets and OGLE-2016-BLG-0596Lb (green pentagon).  Planets are colored
by path to detection: detected and characterized by followup observations
(blue), detected by survey but characterized by followup (magenta),
and detected and characterized by surveys (green).  Their shapes indicate
the principal caustic feature giving rise the anomaly: planetary (circles),
central (squares), and resonant (triangles).  Planets suffering from the
close/wide degeneracy are shown by two open symbols, whereas those for
which this degeneracy is resolved are shown by a single solid symbol.  
By this system, OGLE-2016-BLG-0596Lb should be a solid green triangle.
}
\label{fig:seven}
\end{figure}

\section{Discussion}

OGLE-2016-BLG-0596Lb is a very high mass-ratio $(q=0.0117)$ planet that
lies projected very close to the Einstein ring $(s=1.075)$, which consequently
generated a huge resonant caustic that required 16 days for the source
to traverse. The underlying event was of quite high magnification
$(A_{\rm point-lens}\sim 100)$, which led to pronounced features at peak.
It therefore would seem to be extremely easy to discover.  While the
data set posted on the OGLE web site are adversely affected by the
nearby variable, it is still the case that a free fit to these data
leads to a solution qualitatively similar to the one presented here
(except that it lacks a measurement of $\rho$).  It is therefore striking
that none of the automated programs nor active individual investigators
that query this site noticed this event (or at least they did not
alert the community to what they found as they do for a wide range
of other events, many less interesting).  This indicate the possibility that there
may be many other planets ``hidden in plain sight'' in existing data.
This is also supported by the planet discoveries 
MOA-2008-BLG-397Lb \citep{Suzuki2014},
OGLE-2008-BLG-355Lb \citep{Koshimoto2014}, and 
MOA-2010-BLG-353Lb \citep{Rattenbury2015}, for which the planetary signals
were not noticed during the progress of events.  

These three characteristics, high-magnification (which is usually 
associated with survey+followup rather than survey-only mode),
very high mass ratio, and apparent failure of both machine and
by-eye recognition of the planetary perturbation, prompt us to
address two questions.  First, how do the real (as opposed to
theoretical) planet sensitivities differ between survey-only and
survey+followup modes.  Second, why was this planet discovered
only based on systematic analysis and what does this imply about
the need for such systematic analysis of all events?

\subsection{Summary of Microlens Planet Detections in the Observational
$(s,q)$ Plane}

Many papers contain figures that summarize microlensing planet detections
in the physical plane of planet-mass versus projected separation
(with the latter sometimes normalized by the snow line), 
e.g., Figure 1 of \citet{mb13605}.  And there are many studies that
show plots of {\it planet sensitivity} in the observational $(s,q)$ plane,
(e.g., \citealt{gaudi02,gould10}).  But to our knowledge, there are no
published figures (or even figures shown at conferences) showing the
census of microlensing planet discoveries on this plane.

\begin{figure}
\includegraphics[width=\columnwidth]{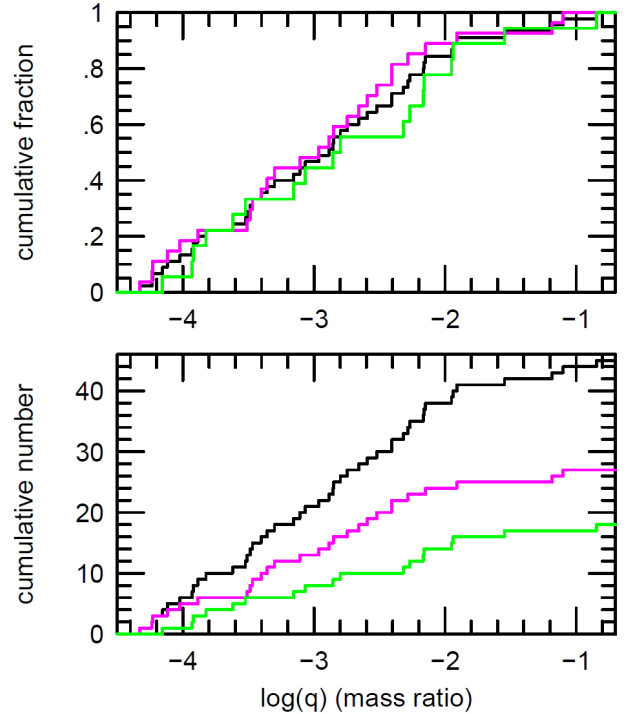}
\caption{
Cumulative microlensing planet detections by log mass ratio 
$\log(q)$, with top normalized and bottom unnormalized.  Green shows
the 18 planets that were detected and characterized by surveys, while
magenta show the 27 planets that required significant followup observations
for detection and/or characterization.  Black is total.  The green and
magenta curves are not statistically distinguishable.
}
\label{fig:eight}
\end{figure}

Figure~\ref{fig:seven}
illustrates the position of OGLE-2016-BLG-0596 (green pentagon) among the
44 previously published planets (or, to the extent we have 
such knowledge, submitted
for publication).  Discovered bodies are defined to be ``planets'' if their
measured or best-estimated mass $m_p<13\,M_{\rm jup}$ and if they are known
to orbit a more massive body\footnote{To facilitate comparison with
future compilations, we list here the 45 planets used to construct
this figure and those that follow. We compress, e.g., OGLE-2003-BLG-235Lb to
OB03235 for compactness and only use ``b,c'' for multiple planets:
OB03235, OB05071, OB05169, OB05390, MB06bin1, OB06109b, OB06109c, MB07192, MB07400,
OB07349, OB07368, MB08310, MB08379, OB08092, OB08355, MB09266, MB09319, MB09387, 
MB10073, MB10328, MB10353, MB10477, MB11028, MB11262, MB11293, MB11322, OB110251,
OB110265, OB120026b, OB120026c, OB120358, OB120406, OB120455, OB120563, OB120724, 
MB13220, MB13605, OB130102, OB130341, OB140124, OB141760, OB150051, OB150954, 
OB150966, OB160596.}.  
Planets are color-coded by discovery method: discovered by followup observations 
(blue), discovered (or discoverable)
in survey-only observations but requiring followup for full-characterization
(magenta), fully (or essentially fully) characterized by survey observations
(green).  The shapes of the symbols indicate the type of caustic that gave rise
to the planetary perturbation: circles, squares, and triangles for
planetary, central, and resonant caustics, respectively.  In many cases,
solutions with $(q,s)$ and $(q,1/s)$ yield almost equally good fits
to the data \citep{griest98}.  In these cases, the two solutions
are shown as open symbols in order to diminish their individual visual
``weight'' relative to the filled symbols used when this degeneracy is
broken.  Hence, OGLE-2016-BLG-0596Lb would be a green filled triangle
if it were not being singled out by making it larger pentagon.

\begin{figure}
\includegraphics[width=\columnwidth]{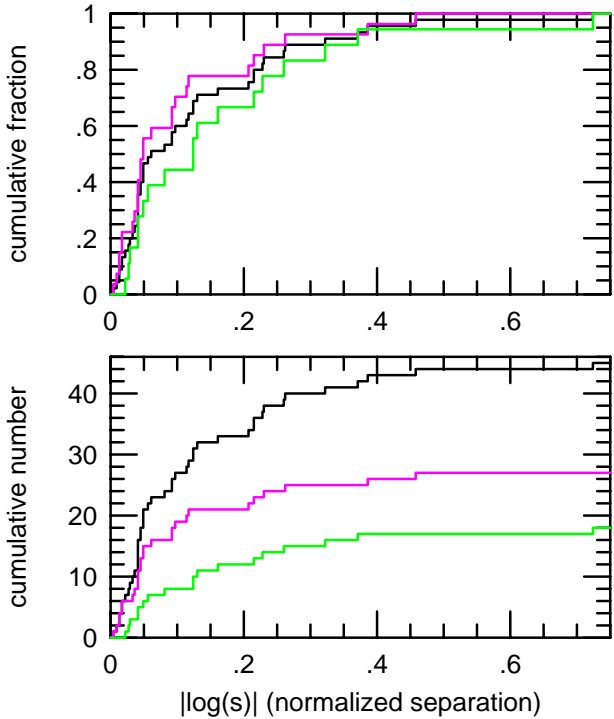}
\caption{
Cumulative microlensing planet detections by 
absolute value of the log projected separation (normalized to $\theta_\e$)
$|\log(s)|$, with top normalized and bottom unnormalized.  Colors are the
same as in Figure~\ref{fig:eight}.  The gap between the green (survey)
and magenta (followup) curves has a 8.5\% probability of being random.
If real, this indicates that followup observation have been relatively
more sensitive to planets near the Einstein ring while surveys are more
sensitive to those further from the Einstein ring.
}
\label{fig:nine}
\end{figure}

The most striking feature of this figure is that, in sharp contrast
to the triangular appearance of high-magnification-event
planet-sensitivity plots (e.g., \citealt{gould10})
and to ``double pronged'' low-magnification sensitivity
plots (e.g., \citealt{gaudi02}), 
this detection plot looks basically like a cross,
with a vertical band of detections near $\log s\sim 0$ and a horizontal
band near $\log(q)\sim -2.5$.  The part of this structure at high mass ratio
$\log(q)>-2$ is easily explained: companions with high mass ratio are,
a priori, most likely stars or brown dwarfs (BDs) and can only be claimed
as ``planets'' if the host mass is known to be low.  This  in turn 
usually requires a measurement of the microlens parallax, which for
ground based observations is much more likely if there is a large caustic
and so $s\sim 1$.

We note that there are 4 planet detections in the region $(\log(s)>+0.15, \log(q)<-3)$, 
while there is no detection in the opposite quadrant $(\log(s)<-0.15, \log(q)<-3)$.
All the 4 planets derive from planetary caustics and 3 of them are pure survey detections:
MOA-2011-BLG-028Lb\footnote{We note that this event's light curve does contain some 
followup data, but it is not essential for characterizing the planet.}, 
OGLE-2008-BLG-092Lb, and MOA-2013-BLG-605Lb \citep{mb11028,ob08092,mb13605}.  
The remaining planet, OGLE-2005-BLG-390Lb \citep{ob05390}, dates from an era when 
followup groups intensively monitored the wings of events, primarily due to the 
paucity of better targets.  Thus we may expect that surveys will gradually fill 
in this quadrant.  The difference in the detection rates between the quadrants 
with $\log(s)>+0.15$ and $\log(s)<-0.15$ can be explained by the difference in 
the size of the planetary caustics with $s<1$ and $s>1$.  In the case of $s>1$, 
there exist a single planetary caustic.  In the case of $s<1$, on the other hand, 
there exist two sets of planetary caustics and each one is smaller than the planetary 
caustic with $s>1$.  As a result, the planetary caustic with $s>1$ has a larger cross 
section and thus higher sensitivity.  Furthermore, smaller caustic size of planets 
with $s<1$ makes planetary signals tend to be heavily affected by finite-source effects, 
which diminish planetary signals, while signals of planets with $s>1$ can survive and 
show up in the wings of light curves.  Actually, all 4 events with planet detections 
via planetary-caustic perturbations are involved with large source stars, i.e.\ giant 
and subgiant stars for which finite-source effects are important.

\begin{figure}
\includegraphics[width=\columnwidth]{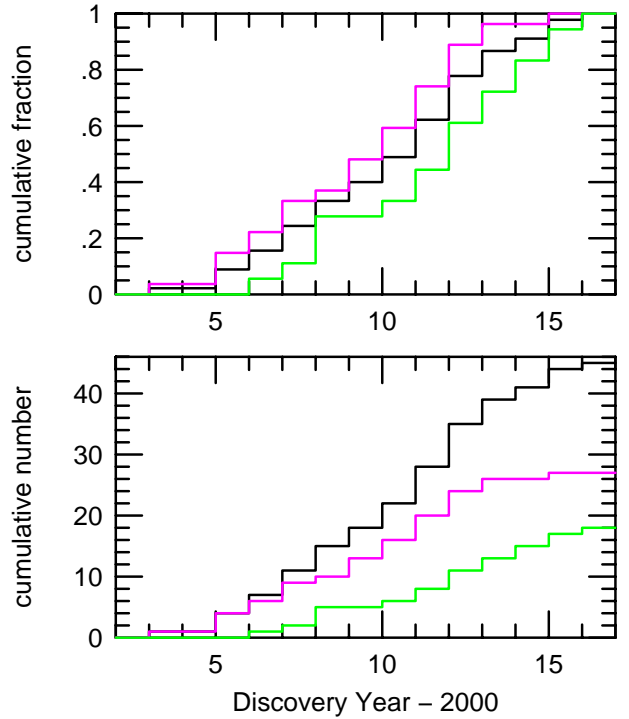}
\caption{
Cumulative microlensing planet detections by year of discovery,
with top normalized and bottom unnormalized.  Colors are the
same as in Figure~\ref{fig:eight}.  Followup discoveries (magenta)
have dropped off dramatically since 2013
}
\label{fig:ten}
\end{figure}

Apart from this quadrant, it is not obvious that surveys are probing
a different part of parameter space from the previously dominant
survey+followup mode.  To further investigate this, we show in 
Figures~\ref{fig:eight} and \ref{fig:nine} the cumulative distributions
of planets by log mass ratio $\log(q)$ and (absolute value of)
log separation $|\log(s)|$.  In this case we distinguish only between
events that could be fully characterized by survey observations (green)
and those that required significant followup (including auto-followup by
surveys).  These distributions generally appear quite similar.
For the mass ratio distribution, the greatest difference (0.259) is at
$\log(q)=-2.319$, which is very typical (Kolmogorov-Smirnoff (KS) probability
40\%).  The greatest difference for the separation distribution
(0.334 at $|\log(s)|=0.124$) has a KS probability of 8.5\%.
This may be indicative of a real difference.  If so, the difference
would be that pure-survey is relatively more efficient at finding
widely separated lenses, which was already hinted at by inspection
of the $(q,s)$ scatter plot.

Finally, in Figure~\ref{fig:ten}, we show cumulative distributions
by year of discovery.  One might expect that with the massive ramp-up
of surveys, survey-only discoveries would move strongly ahead
of survey+followup. This expectation is confirmed in its sign
but not its magnitude by Figure~\ref{fig:ten}.  It shows that in
(2014, 2015, 2016) there have been 
(2,2,1) and (0,1,0) discoveries by survey-only and survey+followup, respectively.
This is certainly not a complete accounting, in part because 2016 has just
begun and in part because historically there has been a considerable
delay in microlensing planet publications for a variety of reasons.
For example, of the 28 planets discovered prior to 2012, the
number with delays (publication year minus discovery year) of 
$(0,1,\ldots,9)$ years
was $N=(1,5,9,5,1,2,4,0,0,1)$.  In the history of microlensing, there has been
only one planet published during the discovery year, OGLE-2005-BLG-071Lb
\citep{ob05071}.  Hence, we will only get a full picture of this
transition after a few years.

\subsection{Challenges to the By-Eye and By-Machine discovery of
OGLE-2016-BLG-0596}

There are three interrelated reasons why OGLE-2016-BLG-0596 may
have escaped notice as a potentially planetary event until the
KMTNet data for this event were examined (for reasons unrelated to
any apparent anomaly).  First, it is relatively
faint at peak.  Second, it has a variable baseline.  Third,
it was not announced as a microlensing event until one day after the
peak.

As a general rule, high-magnification events are singled out for
intensive followup observations only if they are still rising.
When such intensive observations would have been 
conducted,
they
would have immediately revealed the anomalous nature of the event,
probably triggering additional observations.  This is how many
of the planets discovered by $\mu$FUN were found.  While $\mu$FUN 
itself is now semi-dormant,
its protocols are directly relevant here because what is of interest
is whether there is prima facie evidence for a population of missed
planets during past years, during most of which $\mu$FUN was active.

Now, in fact, OGLE-2016-BLG-0596 met the criteria for an OGLE alert
24 hours previously, but no alert was issued because of caution
due to the variable baseline.
Nevertheless, even if such an alert had been issued, it would not
have triggered any followup observations because (due to the anomaly)
the event would have appeared to have already peaked at that time.

Finally, the variability of the baseline may have influenced modelers
and followup groups to discount the evident irregularities in the
light curve near peak as being due to data artifacts.  This could
have been exacerbated by the faintness of the event, which increases both
the formal error bars and the probability of centroiding errors (hence
irregular photometry) due to bright blends.  Both of these effects
reduce the confidence of modelers that apparent anomalies in
online ``quick look'' photometry are due to physical effects.

It is nevertheless a fact that when the original OGLE data are modeled,
they show a clear signal for a massive planet or low-mass BD, which
would trigger a re-reduction of the data, such as the one we report here.

We therefore conclude that while OGLE-2016-BLG-0596 has some 
near-unique features that increased the difficulty of recognizing
it as a planetary event, such recognition was clearly feasible.
Hence, we do indeed regard this event as prima facie evidence for
more such events in archival data, particularly OGLE-IV data 2010-2015.

\acknowledgments 
The OGLE project has received funding from the National Science Centre,
Poland, grant MAESTRO 2014/14/A/ST9/00121 to AU.
Work by C.H.\ was supported by Creative Research Initiative
Program (2009-0081561) of National Research Foundation of Korea.
The OGLE Team thanks Profs. M.~Kubiak and G.~Pietrzy{\'n}ski, former
members of the OGLE team, for their contribution to the collection of
the OGLE photometric data over the past years.
WZ and AG were supported by NSF grant AST-1516842.
Work by JCY was performed under contract with the California Institute of Technology 
(Caltech)/Jet Propulsion Laboratory (JPL) funded by NASA
through the Sagan Fellowship Program executed by the
NASA Exoplanet Science Institute.
This research has made the telescopes
of KMTNet operated by the Korea Astronomy and Space Science
Institute (KASI).


\begin{thebibliography}{99}
\bibitem[Alard \& Lupton(1998)]{alard98} Alard, C. \& Lupton, R.~H.\ 1998, 
         \apj, 503, 325
\bibitem[Beaulieu et al.(2006)] {ob05390} Beaulieu, J.-P. Bennett, D.~P., 
         Fouqu\'e, P., et al.\ 2006, Nature, 439, 437
\bibitem[Bennett et al.(2008)]{mb07192}  Bennett, D.~P., Bond, I.A., 
         Udalski, A., et al.\ 2008, \apj, 684, 663
\bibitem[Bennett et al.(2012)]{moabin1}  Bennett, D.~P., Sumi, T., 
         Bond, I.~A., et al.\ 2012, \apj, 757, 119
\bibitem[Bensby et al.(2013)]{bensby13} Bensby, T. Yee, J.~C., Feltzing, S., 
         et al.\ 2013, \aap, 549A, 147
\bibitem[Bessell \& Brett(1988)]{bb88} Bessell, M.~S., \& Brett, J.~M.\ 1988, 
         \pasp, 100, 1134
\bibitem[Bond et al.(2004)]{ob03235} Bond, I.~A., Udalski, A., Jaroszy\'nski, M, 
         et al.\ 2004, \apj, 606, L155
\bibitem[Gaudi et al.(2002)]{gaudi02} Gaudi, B.~S., Albrow, M.~D., An, J.\ 2002, 
         \apj, 566, 463
\bibitem[Gould \& Loeb(1992)]{gouldloeb} Gould, A. \& Loeb, A.\ 1992, \apj, 396, 104
\bibitem[Gould et al.(2010)]{gould10} Gould, A., Dong, S., Gaudi, B.~S., et al.\ 2010, 
         \apj, 720, 1073
\bibitem[Griest \& Safizadeh(1998)]{griest98} Griest, K.\ \& Safizadeh, N.\ 1998, 
         \apj, 500, 37
\bibitem[Hirao et al.(2016)]{Hirao2016} Hirao, Y., Udalski, A., Sumi, T., et al.\ 2016, 
         \apj, 824, 139
\bibitem[Kervella et al.(2004)]{kervella04} Kervella, P., Th{\'e}venin, F., Di Folco, E., 
         \& S{\'e}gransan, D.\ 2004, \aap, 426, 297
\bibitem[Kim et al.(2016)]{kmtnet} Kim, S.-L., Lee, C.-U., Park, B.-G., et al.\ 2016, 
         JKAS, 49, 37
\bibitem[Koshimoto et al.(2014)]{Koshimoto2014} Koshimoto, N., Udalski, A., 
         Sumi, T., et al.\ 2014, \apj, 788, 128
\bibitem[Nataf et al.(2013)]{nataf13} Nataf, D.~M., Gould, A., Fouqu\'e, P., 
         et al.\ 2013, \apj, 769, 88
\bibitem[Poleski et al.(2014a)]{ob08092} Poleski, R., Skowron, J., Udalski, A., 
         et al.\ 2014a, \apj, 755, 42
\bibitem[Poleski et al.(2014b)]{ob120406} Poleski, R., Udalski, A., Dong, S., 
         et al.\ 2014b, \apj, 782, 47
\bibitem[Rattenbury et al.(2015)]{Rattenbury2015} Rattenbury, N.~J., Bennett, D.~P., 
         Sumi, T., et al.\ 2015, \mnras, 454, 946
\bibitem[Schechter et al.(1993)]{dophot} Schechter, P.~L., Mateo, M., \& 
         Saha, A.\ 1993, \pasp, 105, 1342
\bibitem[Shin et al.(2016)]{ob150954} Shin, I.-G., Ryu, Y.~H, Udalski, A., et al.\ 2016, 
         JKAS, 49, 73
\bibitem[Shvartzvald et al.(2014)]{mb11322} Shvartzvald, Y., Maoz, D., Kaspi, S., 
         et al.\ 2014, \mnras, 439, 604
\bibitem[Skowron et al.(2011)]{mb11028} Skowron, J., Udalski, A., Poleski, R., 
         et al.\ 2016, \apj, 820, 4
\bibitem[Skowron et al.(2016)]{skowron16} Skowron, J., Udalski, A., Koz{\l}owski, S., 
         et al.\ 2016, Acta Astron., 66, 1
\bibitem[Sumi et al.(2016)]{mb13605} Sumi, T., Udalski, A., Bennett, D.~P., 
         et al.\ 2016 \apj, in press arXiv:1512.00134
\bibitem[Suzuki at al.(2014)]{Suzuki2014} Suzuki, D., Udalski, A., Sumi, T., 
         at al.\ 2014, \apj, 780, 123	
\bibitem[Udalski(2003)]{ews2} Udalski, A.\ 2003, Acta Astron., 53, 291
\bibitem[Udalski et al.(1994)]{ews1} Udalski, A.,Szymanski, M., Kaluzny, J., 
         Kubiak, M., Mateo, M.,  Krzeminski, W., \& Paczy\'nski, B.\ 1994, 
         Acta Astron., 44, 317
\bibitem[Udalski et al.(2005)]{ob05071} Udalski, A., Jaroszy\'nski, M., Paczy\'nski, B, 
         et al.\ 2005, \apj, 628, L109.
\bibitem[Udalski et al.(2015b)]{ogleiv} Udalski, A., Szyma\'nski, M.K. \& Szyma\'nski, 
         G.\ 2015b, Acta Astronom., 65, 1
\bibitem[Yee et al.(2012)]{mb11293} Yee, J.C., Shvartzvald, Y., Gal-Yam, A., 
         et al.\ 2012, \apj, 755, 102
\bibitem[Yoo et al.(2004)]{ob03262} Yoo, J., DePoy, D.~L., Gal-Yam, A., et al., 
         2004, \apj, 603, 139
\bibitem[Zhu et al.(2014)]{Zhu:2014} Zhu, W., Penny, M., Mao, S., Gould, A., \& 
         Gendron, R.\ 2014, \apj, 788, 73




\end{thebibliography}
\end{document}